\documentclass[reprint,10pt,onecolumn,a4paper,superscriptaddress,longbibliography,
amsmath,amssymb]{revtex4-2}

\usepackage{lineno}
\usepackage{mathtools}
\usepackage{amsmath}
\usepackage{graphicx}
\usepackage{booktabs,multirow}
\usepackage{braket}
\usepackage{textcomp}
\usepackage{yfonts}
\usepackage{comment}
\usepackage{array}
\usepackage[table]{xcolor}

\usepackage[normalem]{ulem}

\usepackage{geometry}
\usepackage[%
  bookmarks=true,
  colorlinks,
  linkcolor=blue,
  urlcolor=blue,
  citecolor=blue,
  plainpages=false,
  pdfpagelabels,
  final,
  breaklinks=true
]{hyperref}
\usepackage{orcidlink}
 
\begin{document}
\newcommand{\pd}[2]{\frac{\partial #1}{\partial #2}} 
\newcommand{\td}[2]{\frac{d #1}{d #2}} 

\newcommand{\bs}{\boldsymbol}
\newcommand{\bt}{\textbf}
\newcommand{\sech}{\text{sech}}
\newcommand{\erfc}{\text{erfc}}
\newcommand{\bse}{\begin{subequations}}
\newcommand{\ese}{\end{subequations}}
\newcommand{\rect}{\text{rect}}
\newcommand{\sgn}{\text{sgn}}
\newcommand{\im}{\text{i}}
\newcommand{\ud}[0]{\mathrm{d}}
\newcommand{\norm}[1]{\left\lVert#1\right\rVert}
\newcommand{\op}{\widehat}

\graphicspath{{figs/},{../figs/}} 
\allowdisplaybreaks

\title{Binary holograms for shaping light with digital micromirror devices}

\author{R. Guti\'{e}rrez-Cuevas~\orcidlink{0000-0002-3451-6684}}
\email{rodrigo.gutierrez-cuevas@espci.fr}
\affiliation{Institut Langevin, ESPCI Paris, Université PSL, CNRS, 
75005 Paris, France}
\author{S. M. Popoff~\orcidlink{0000-0002-7199-9814}}
\email{sebastien.popoff@espci.fr}
\affiliation{Institut Langevin, ESPCI Paris, Université PSL, CNRS, 
75005 Paris, France}

\date{\today}

\begin{abstract}
Digital micromirror devices are a popular 
type of spatial light modulators for wavefront shaping applications. 
While they offer several advantages when compared to liquid crystal modulators, such as polarization insensitivity and rapid-switching, they only provide a binary amplitude modulation. 
Despite this restriction, it is possible to use binary holograms to modulate both the amplitude and phase of the incoming light, thus allowing the creation of complex light fields. 
Here, a didactic exploration of various types of binary holograms is presented.
A particular emphasis is placed on the fact that the finite number of pixels coupled with the binary modulation limits the number of complex values that can be encoded into the holograms. 
This entails an inevitable trade-off between the number of complex values that can be modulated with the hologram and the number of independent degrees of freedom available to shape light, both of which impact the quality of the shaped field. 
Nonetheless, it is shown that by appropriately choosing the type of hologram and its parameters, it is possible to find a suitable compromise that allows shaping a wide range of complex fields with high accuracy.
In particular, it is shown that choosing the appropriate alignment between the hologram and the micromirror array allows for maximizing the number of complex values.
Likewise, the implications of the type of hologram and its parameters on the diffraction efficiency are also considered. 
\end{abstract}

\maketitle


\section{Introduction}



The ability to control light's various degrees of freedom has enabled many new applications as well as improved existing ones across many fields 
~\cite{piccardo2021roadmap,andrews2008structured,rubinszteindunlop2017roadmap,taylor2015enhanced,gigan2022roadmap}. 
For instance, it allowed shaping the intensity distribution of optical fields for imagining samples with nanometric resolution~\cite{hell1994breaking,backer2014extending,jouchet2021nanometric,vicidomini2018sted}, and creating optical tweezers for manipulating small particles~\cite{ashkin1986observation,neuman2004optical,callegari2015computational,volpe2023roadmap,gutierrezcuevas2018lorenz}.
Likewise, this control enabled the creation of fields with non-trivial phase structures such as optical vortices which have been used for quantum information applications~\cite{mirhosseini2015high,malik2016multi,krenn2017orbital}, among many others~\cite{yao2011orbital,dennis2009chapter}. 
A more recent use of wavefront shaping is for studying and controlling light's propagation through complex media~\cite{rotter2017light,gigan2022roadmap}.
In particular, it allows measuring parts of their scattering matrix which establishes a linear relationship between a set of inputs and output modes, and provides a high-level of control over the wave propagation~\cite{popoff2011controlling,popoff2010image,popoff2010measuring}. 
The knowledge of this matrix can then be used for focusing or imaging through highly-scattering media~\cite{vellekoop2007focusing,ambichl2017focusing,gigan2022imaging}, creating highly-precise sensing devices~\cite{bouchet2021maximum,gutierrez2023crashing}, and even avoiding undesirable effects~\cite{carpenter2015observation,ambichl2017super,matthes2021learning}. 
The huge success of structured light has also led to many works seeking to understand the limits within which light can be shaped. 
This endeavor has resulted in the creation of various nontrivial topological structures with optical field, such as knots~\cite{dennis2010isolated,pisanty2019knotting,larocque2018reconstructing}, Möbius strips \cite{freund2005cones,bauer2015observation,bliokh2019geometric}, and skyrmions \cite{du2019deep,tsesses2018optical,gutierrezcuevas2021optical}.

Over the years, there have been several key technological advances that have allowed the creation and use of structured light fields. 
Holograms were initially used to encode the spatial distribution of a target field into a photographic film \cite{gabor1948new,gabor1949microscopy,leith1962reconstructed}. 
However, the printing process relied on exposing the film to the interference pattern created by the superposition of a reference field with the desired target.
Therefore, these off-axis holograms required the existence of the target field before it could be encoded into the hologram.  
With the inception of computer generated holograms, this restriction was lifted \cite{brown1966complex,fienup1974new,haskell1973computer,lee1979binary,lee1974binary,burch1967computer,lee1978iii}. 
Any target field could now be encoded into a phase or amplitude hologram and subsequently generated by illuminating the hologram thus opening the door for shaping complex light fields. 
More recently, spatial light modulators (SLMs) were developed as reconfigurable devices in which one could encode any hologram with a computer. 
Two popular types of SLMs are: liquid crystal SLMs which provide a continuous modulation of the phase \cite{konforti1988phase,barnes1989phase,arrizon2007pixelated,rosalesguzman2017how} and digital micro-mirror devices (DMDs) which provide a binary (0 or 1) modulation of the amplitude \cite{hornbeck1997digital,dudley2003emerging,scholes2019structured,ren2015tailoring}. 
Despite their own limitations, they can both be used to shape the amplitude and phase of optical fields using the same computer generated holograms that were proposed more than half a century ago \cite{lee1978iii}.



DMDs present several advantages over liquid-crystal SLMs, they are more affordable, allow for faster switching (on the order of tens of kHz), and are polarization insensitive~\cite{akbulut2011focusing,conkey2012high}. 
Despite their binary amplitude modulation, there are binary holograms that allow achieving a complex (amplitude and phase) modulation but at the expense of reducing the number of degrees of freedom and lowering the transmittance. 
There have been other works that have reviewed various aspects of DMDs and their light shaping capabilities. 
However, they have been written to provide an overview of their technical aspects, how to use them to shape light, and their main applications~\cite{dudley2003emerging,park2015properties,ren2015tailoring,scholes2019structured,popoff2023DMDtutorial}.
Here, a different approach is taken by focusing solely on how to best encode a complex valued optical field onto a binary hologram created on a DMDs. 
Therefore, special care is taken to consider the implications of the discrete, and often Cartesian, layout of micro-mirrors on the complex values that can be encoded in the hologram.
In particular, it is explicitly shown how this leads to a discretization of the attainable complex values and how their number can be drastically increased by appropriately choosing the angle between the hologram structure and the axes of the micro-mirror array. 
It is also shown how there is a compromise between the number of complex values that can be used to shape light, the diffraction efficiency, and the final resolution of the shaped field which is equivalent to the number of degrees of freedom in the SLM. 
All the code used to generate the results presented here  
along with examples of use can be accessed at the dedicated GitHub repository \cite{gutierrezcuevaspyDMDholo}. 

\section{Gray-scale amplitude modulated holograms}

\subsection{Interference-generated holograms}

\begin{figure}
  \centering
  \includegraphics[width=.95\linewidth]{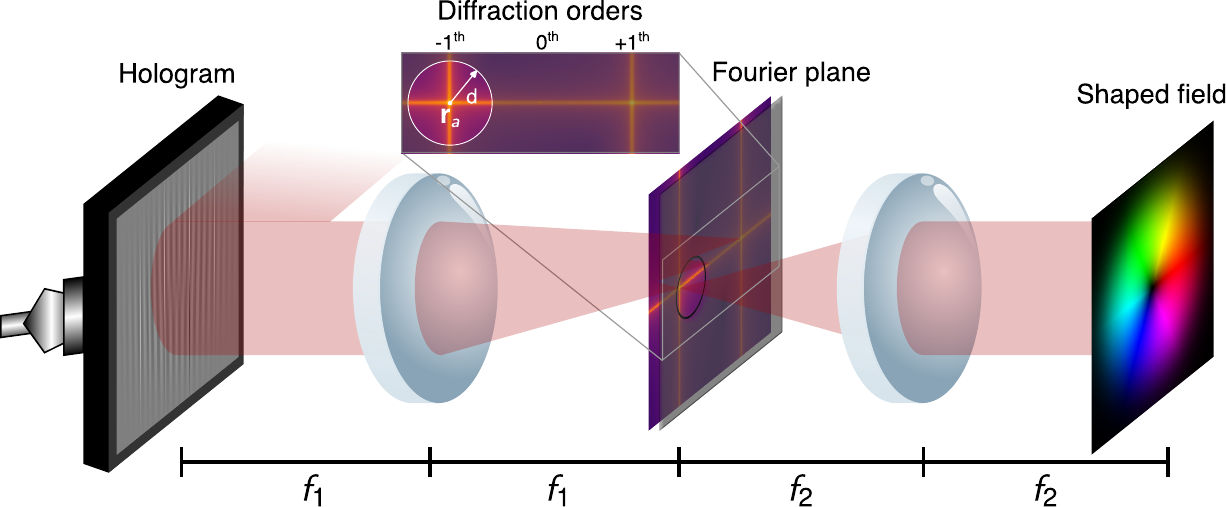}
  \caption{\label{fig:setup} 
  Experimental setup for shaping optical beams with amplitude modulated holograms. An input plane wave illuminates an amplitude hologram. 
  The diffracted light is then Fourier transformed by a lens. 
  At the Fourier plane, the various diffraction orders are spatially separated and the one containing the information about the target field can be filtered by placing a small aperture. 
  Note that the optical axis is chosen to be aligned with the filtered diffraction order to avoid a residual phase ramp in the shaped field. 
  Finally, the filtered diffraction order is Fourier transformed in order to create the desired output shaped field.}
\end{figure}

Amplitude modulated holograms find their roots in the interferograms that were originally recorded on photographic film by interfering two fields~\cite{gabor1948new,gabor1949microscopy,leith1962reconstructed,lee1978iii}. 
In these holograms, the spatial structure of a given target (or object) field,  $E_t (\bt r)$, 
is recorded by making it interfere with a tilted reference planewave, $E_r(\bt r)=R \exp(\im 2\pi \bs \nu \cdot \bt r)$, whose direction is determined by the carrier frequency vector $\bs \nu$. 
After the recording process, the transmittance function of the photographic plate follows the same spatial dependence as that of the intensity distribution produced by the interference between the two waves.
By expressing the target field in terms of its amplitude $A(\bt r)$ and phase $\phi(\bt r)$, so that $E_t (\bt r)=A(\bt r) \exp [\im \phi(\bt r)] $, the transmittance function of the hologram can be written as 
\begin{align}
  \label{eq:oaiholo}
  T(\bt r) = & |E_r(\bt r)+E_t(\bt r)|^2 = R^2 +A^2(\bt r) + 2RA(\bt r) \cos\left[2\pi \bs \nu \cdot \bt r -\phi(\bt r)\right] .
\end{align} 
As shown in Fig.~\ref{fig:setup}, the resulting hologram takes the form of a spatially varying grating where $\bs \nu$ defines the orientation of the fringes. 
In principle, this transmittance function can take any value between $0$ and $1$ and is thus referred to as a grayscale hologram.  
Moreover, from the transmittance function, it can be seen that the phase information of the target field is encoded into the position of the fringes while the amplitude information is encoded on the peak-to-valley value of fringes. 

In order to understand how to retrieve the target field from the hologram, it is useful to rewrite the transmittance function in terms of a Fourier series.
Assuming that the spatial variations of the target field are slower than the amplitude modulations controlled by $\bs \nu$, then the hologram is locally a two-dimensional periodic structure thus forming a Bravais lattice, just like the ones found in crystals \cite{ashcroft2022solid}. 
This two-dimensional lattice is fully determined by the two primitive vectors, $\bt a_1$ and $\bt a_2$, defining the unit cell and which need not be orthogonal. 
Given this periodic structure, it is then possible to rewrite the transmittance function as a Fourier series,
\begin{align} \label{eq:fs}
  T(\bt r) = \sum_{m_1=-\infty}^{\infty}\sum_{m_2=-\infty}^{\infty} T_{m_1 m_2}
  e^{2\pi \im m_1 \bt b_1 \cdot \bt r}
  e^{2\pi \im m_2 \bt b_2 \cdot \bt r}
\end{align}
where the spatial dependence of the target field is encoded into the Fourier coefficients $T_{m_1m_2}$. 
Here, $\bt b_1$ and $\bt b_2$ are the reciprocal lattice vectors which are defined via $\bt a_i \cdot \bt b_j = \delta_{ij}$ with $\delta_{ij}$ being the Kronecker delta. 
Given that the spatial variation of the target field are assumed to be much slower than those controlled by $\bs \nu$, the spatial dependence of the field will be omitted in the remainder of this work.
In many cases, holograms are invariant along one direction thus defining a one-dimensional lattice with a single primitive vector pointing in the same direction as the reciprocal vector and a single sum in Eq.~(\ref{eq:fs}).

When the hologram is subsequently illuminated by a plane wave, assumed to be propagating along the optical axis for simplicity, the field coming out of it is proportional to its transmittance function and is thus composed of many waves traveling along different directions defined by each term in the Fourier series with a nonzero Fourier coefficient. 
These nonvanishing terms  correspond to the diffraction orders of the grating structure encoded into the hologram \cite{lee1978iii,goodman2005introduction}. 
Each diffraction order carries a fraction of the power incident on the grating. This fraction defines the diffraction efficiency of a given order which is determined by the corresponding Fourier coefficient via
\begin{align}
  \eta_d^{(m_1,m_2)} = |T_{m_1 m_2}|^2.
\end{align} 


For the particular case of the interference-generated hologram given in Eq.~(\ref{eq:oaiholo}), the Fourier series takes the following simple form,
\begin{align} \label{eq:origholo}
  T(\bt r) = &  R^2 +A^2 + RA e^{-\im \phi}
        e^{ 2\pi \im \bs \nu \cdot \bt r}
         + RA e^{\im  \phi}e^{- 2\pi\im  \bs \nu \cdot \bt r},
\end{align}
where only three Fourier components are nonzero, and there is only one primitive vector whose reciprocal vector is equal to the carrier frequency vector, that is $\bt b = \bs \nu$.  
Each diffraction order has a particular structure encoded into the corresponding Fourier coefficients. 
In particular, the Fourier coefficient for the diffraction order $m=-1$ is proportional to the target field that was encoded. 
Therefore, in order to be able to shape the input light into the target field, it is necessary to separate this diffraction order from the others. 
This is done via the $4f$ system shown in Fig.~\ref{fig:setup}. 
First, the field produced by illuminating the hologram is propagated through a $2f$ system in order to obtain the Fourier transform of the transmittance function. 
In the Fourier plane, all the diffraction orders are separated spatially so that the one containing the target field information can be filtered by placing a small aperture.
Then, the filtered diffraction order is propagated through another 2-$f$ system in order to perform a second Fourier transform and thus obtain a shaped field which ideally should be equal to the target up to a homothetic transformation. 

When encoding a complex field into a hologram, one needs to fix a value for the carrier frequency vector $\bs \nu$. The norm of $\bs \nu$ needs to be large enough to be able to sample the target field correctly and to avoid overlapping information between neighboring diffraction orders. 
Both are ensured by taking $\norm{\bs \nu}$ at least larger than twice the bandwidth of the target field in accordance to the Nyquist sampling theorem~\cite{goodman2005introduction}.
Or, vice versa, if $\bs \nu$ is fixed then only fields with a bandwidth smaller than $\norm{\bs \nu}/2$ can be shaped appropriately.
Once $\bs \nu$ has been fixed, the location of the desired diffraction order and thus of the aperture is given by $\bt r_a =-\lambda f_1 \bs \nu$ with $\lambda$ being the wavelength of the incident light and $f_1$ the focal length of the first lens in the $4f$ system.
The size of the aperture, $d$, needs to satisfy two conditions. 
It needs to be large enough so that the full bandwidth of the target field can propagate through but small enough so that no light from the surrounding diffraction orders can pass. 
The optimal choice of these parameters depend on the type of target fields one wishes to produce and the medium in which the hologram is encoded as will be seen in the following sections.
Lastly, given that the shaped field is the desired output, it makes sense to align the optical axis of the $4f$ system with respect to its propagation direction as depicted in Fig.~\ref{fig:setup}. 
By doing this, the filtered diffraction order lies on the optical axis and is centered with respect to the second lens, thus removing the phase tilt from the shaped field. 
Otherwise, the output shaped field will come at an angle which needs to be taken into account.

\subsection{Computer-generated holograms}

With the advent of computer generated holograms~\cite{brown1966complex,fienup1974new,haskell1973computer,lee1979binary,lee1974binary,burch1967computer,lee1978iii}, it was soon realized that there was no need to restrict oneself to holograms of the form given by Eq.~\eqref{eq:oaiholo}. 
The interference-generated hologram could then be adapted by dropping the unnecessary parts coming from the reference wave and simply use the computer to generate the following transmittance function~\cite{burch1967computer,lee1978iii}, 
\begin{align}
  \label{eq:oaholo}
  T(\bt r)= & \frac 1 2 + \frac 1 2 A \cos\left(2\pi  \bs \nu \cdot \bt r -\phi\right),
\end{align}
where it is assumed that the amplitude $A$ lies between zero and one. This assumption will be made in the remainder of this work. 
It can be easily verified that this off-axis hologram shapes light in the same way as the interference-generated hologram presented in the previous section, that is, by producing three diffraction orders with the $m_1=-1$ one containing the desired target field which is produced with a diffraction efficiency of 6.25\%. 

There are, however, other less obvious alternatives. For example, the transmittance function based on the grayscale Lee sampling hologram~\cite{lee1970sampled,lee1978iii}, 
\begin{align}
  T(\bt r)= & \frac{A}{2}\left[ \cos\left(2\pi  \bs \nu \cdot \bt r -\phi\right) 
   + |\cos \left(2\pi  \bs \nu \cdot \bt r -\phi\right)|\right],
\end{align}
also generates the desired object field. In order to verify that this is indeed the case, the transmittance function is expanded into a Fourier series,
\begin{align}
  T(\bt r)= & \frac{A}{4} \left[e^{-\im \phi} e^{\im 2\pi  \bs \nu \cdot \bt r }+ e^{\im \phi} e^{-\im 2\pi  \bs \nu \cdot \bt r } + \sum_{n=-\infty}^{\infty}c_n e^{-2\im n \phi }e^{\im 4n\pi \bs \nu \cdot \bt r } \right],
\end{align}
where 
\begin{align}
  c_n = \frac{(-1)^{n+1}}{\pi}\frac{4}{4n^2 - 1},
\end{align}
and $m=2n$ with $m$ being the index of the Fourier series as in Eq.~(\ref{eq:fs}).
In this case the hologram produces an infinity of diffraction orders, with the same primitive vector as before. 
Here, again the target field is encoded into the $m=-1$ order, which has the same diffraction efficiency as the off-axis hologram. 

\begin{figure}
  \centering
  \includegraphics[scale=.75]{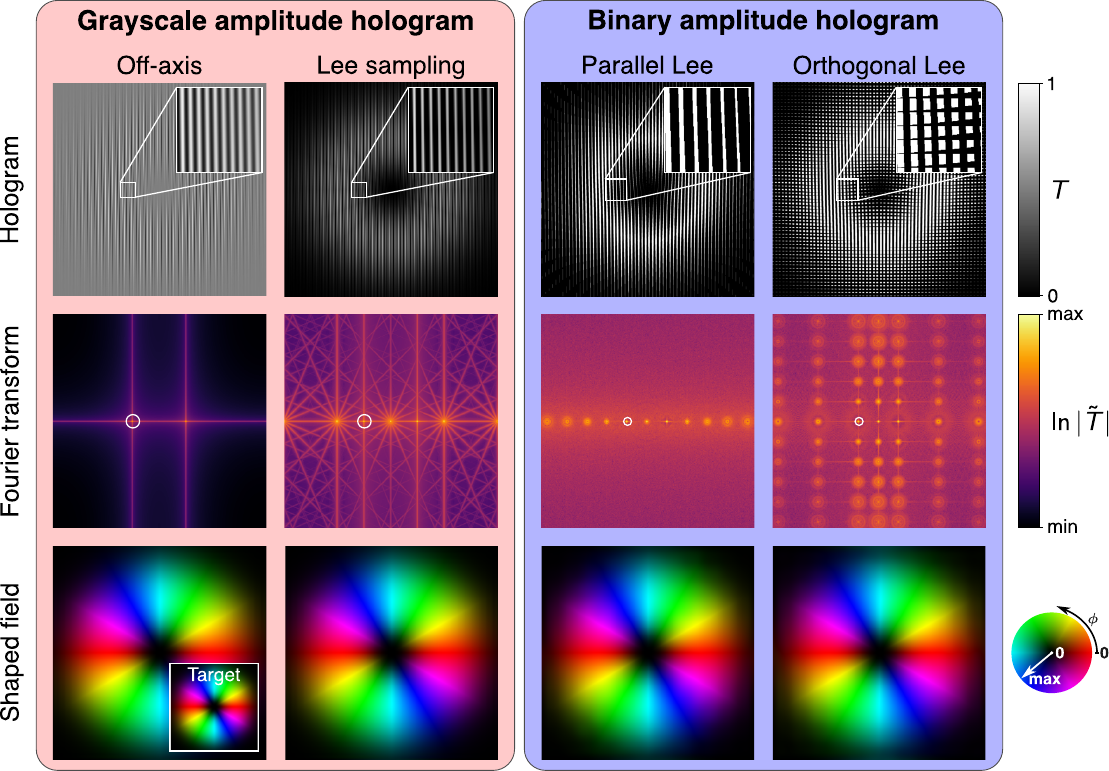}
  \caption{\label{fig:ampcomp}
  Comparison between the amplitude grayscale (first column) off-axis and (second column) sampling Lee holograms, and the binary (third column) parallel and (fourth column) orthogonal Lee holograms for shaping light into a Laguerre-Gauss (LG) beam with a phase vortex of order two. (first row) The transmittance function of the hologram where the inset provides a closer view at the underlying amplitude modulation. (second row) The amplitude of the Fourier transform of the corresponding hologram in log scale where the various diffraction orders are clearly visible. (third row) The resulting shaped field, which for all cases is indistinguishable from the target shown in the inset of the first shaped field.}
\end{figure}

The diffraction order containing the target field information for both holograms can be filtered out using the same setup shown in Fig.~\ref{fig:setup}. 
Figure \ref{fig:ampcomp} shows an example of using these two holograms to shape the input light into a Laguerre-Gauss (LG) beam with a topological charge equal to two \cite{siegman1986lasers,gutierrezcuevas2019generalized,gutierrezcuevas2020modal}. 
Both the off-axis and Lee sampling holograms encode the information about the target field similarly. 
The phase information is encoded into the position of the fringes that make the grating, and the amplitude into the peak-to-valley value of the fringes. 
However, a clear difference between the two can be appreciated, particularly at the Fourier plane where we can appreciate the various diffraction orders. 
Nonetheless, both holograms produce a satisfactory shaped field which is indistinguishable from the target.

\section{Binary amplitude modulated holograms}

In the previous section, it was assumed that the transmittance function can take any value between $0$ and $1$.
However, as was mentioned in the introduction, this is not the case for DMDs which only provide a binary amplitude modulation. 
Even thought the use of DMDs to shape light is relatively recent, the encoding of binary amplitude modulated holograms for shaping light was thoroughly addressed more than half a century ago \cite{brown1966complex,haskell1973computer,lee1974binary,lee1978iii,lee1979binary}.
This was due to the fact that nonlinearities in the printing process made the printing of binary holograms more accurate than grayscale ones \cite{lee1978iii,lee1979binary}. 
The first attempts at encoding a target field into a binary amplitude hologram were directly based on trying to adapt the off-axis gray-scale hologram in Eq.~(\ref{eq:oaholo}) by simply setting all values above the mean to be one and the rest to be zero~\cite{bryngdahl1968interferograms}. 
This leads to a transmittance function of the form
\begin{align}
  \label{eq:phaseholo}
  T(\bt r) = & \frac 1 2 + \frac 1 2 \sgn [\cos\left(2\pi\bs \nu \cdot \bt r -\phi\right)], 
  \end{align}
where
  \begin{align}
  \sgn(x) = 
  \begin{cases}
  1, & x>0 \\
  -1, & \text{otherwise}
  \end{cases}
\end{align}
is the sign function. 
Here, the phase information is still encoded into the position of the fringes that make the holograms. 
However, the information about the amplitude distribution of the target field is lost. 
Nevertheless, it is possible to modify this first approximation in order to regain the amplitude encoding.
There are two main types of holograms that accomplish this, which unfortunately go by the same name of Lee holograms~\cite{brown1966complex,lee1978iii,lee1979binary,lee1974binary}. In order to distinguish them, an adjective will be added to their name based on how the amplitude is encoded into them.


\subsection{Parallel Lee hologram}

The first one is the \emph{parallel} Lee hologram which is obtained when one tries to fix the hologram in Eq.~(\ref{eq:phaseholo}) by encoding the amplitude using a different threshold than the mean value~\cite{brown1966complex,lee1978iii,lee1979binary,lee1974binary}.
This leads to the following transmittance function
\begin{align}
  \label{eq:parlee}
  T(\bt r) = & \frac 1 2 + \frac 1 2 \sgn [\cos\left(2\pi\bs \nu \cdot \bt r -\phi\right) - \cos \left(\pi q \right)],
\end{align}
where $q$ controls the threshold.
To better understand the effect of $q$ and its relation to $A$, it is worth rewriting the transmittance function in the following manner,
\begin{align}
  T(\bt r) = \sum_{n=-\infty}^{\infty} \rect \left(\frac{\bs \nu \cdot \bt r -\phi/2\pi  -n}{q}\right), 
\end{align}
where
\begin{align}
  \rect(x) = 
  \begin{cases}
  1, & |x|<1/2 \\
  0, & \text{otherwise}
  \end{cases}.
\end{align}
Here, it can be seen that the hologram is composed of binary fringes whose duty cycle, or width, is controlled by $q$. 
This change in width can be exploited to encode the amplitude variations of the target field $A$. 
This can be seen by rewriting the transmittance function as a Fourier series given by
\begin{align}
  T(\bt r) = \sum_{m=-\infty}^{\infty} \frac{\sin ( \pi m q)}{\pi m}e^{-\im m  \phi}e^{ 2\pi \im m \bs \nu \cdot \bt r}. 
\end{align}
Here again, there is a single primitive vector whose reciprocal is equal to $\bs \nu$, and there are an infinite number of diffraction orders which are uniformly spaced. 
The target field is encoded into the Fourier coefficient for the $m=-1$ order which is given by 
\begin{align} \label{eq:parleecoef}
  T_{-1} = \frac{A}{\pi} e^{\im \phi}, \qquad \text{with} \qquad 
  A = \sin \pi q,
\end{align}
thus establishing a nonlinear relation between $q$ and $A$.
As can be appreciated in the example shown in Fig.~\ref{fig:ampcomp}, the larger the duty cycle is, the larger the amplitude, and vice versa. 
Under appropriate conditions, and neglecting any effects due to the discretization of the pixels,  the resulting shaped field is indistinguishable from the target, just like for the gray-scaled holograms. 
From Eq.~(\ref{eq:parleecoef}), it can also be seen that the diffraction efficiency is equal to $10.1\%$ which is larger than that of the gray scale holograms. 

\subsection{Orthogonal Lee hologram}

Thus far, all the holograms that have been presented encode both the amplitude and phase information along the same direction determined by the carrier frequency $\bs \nu$. 
In contrast, the \emph{orthogonal} Lee hologram decouples these two encodings by using the orthogonal direction to encode the amplitude distribution of the target field~\cite{brown1966complex,lee1978iii,lee1979binary,lee1974binary}. 
This is achieved by considering the following transmittance function
\begin{align}
  \label{eq:ortlee}
  T = & \left(\frac 1 2 + \frac 1 2 \sgn [\cos\left(2\pi \bs \nu \cdot \bt r -\phi\right) 
  ] 
  \right) 
   \left( \frac 1 2 + \frac 1 2 \sgn [\cos\left(2\pi \bs \nu_\perp \cdot \bt r\right) - \cos \left(\pi A\right)] \right),
\end{align}
where $\bs \nu_\perp$ is the carrier frequency for the amplitude modulation, which satisfies $\bs \nu \cdot \bs \nu_\perp =0$. 
Once more, in order to get a better sense of how this hologram encodes the target field's information, the transmittance function can be rewritten in the following manner
\begin{align}
  T(\bt r) =  &\sum_{n_1=-\infty}^{\infty}\sum_{n_2=-\infty}^{\infty} \rect \left( \frac{\bs \nu \cdot \bt r-\phi/2\pi  -n_2}{1/2}\right) 
    \rect \left( \frac{\bs \nu_\perp \cdot \bt r -n_1}{A}\right).
\end{align}
Just like for the parallel Lee hologram, the phase is encoded in the position of the binary fringes along the $\bs \nu$ direction, but their widths are now held constant.  
The amplitude encoding is done in the $\bs \nu_\perp$ direction in which parts of the fringes are set to zero in order to create an array of rectangles of fixed width but whose height is adjusted to the desired amplitude of the target field following a linear relationship. 
To verify that this hologram indeed produces the desired shaped field, we can again decompose it in terms of its Fourier series
\begin{align}
  T = &\sum_{m_1=-\infty}^{\infty}\sum_{m_2=-\infty}^{\infty} \frac{\sin\left( \pi m_1 /2\right)}{\pi m_1}\frac{\sin \left(\pi m_2 A\right)}{\pi m_2}e^{-\im m_1 \phi}
    e^{\im  2\pi (m_1\bs \nu + m_2\bs \nu_\perp) \cdot \bt r}.
\end{align}
In this case, due to the encoding along both axes, there are two primitive vectors and whose reciprocal vectors are $\bt b_1 = \bs \nu$ and $\bt b_2 = \bs \nu_\perp$ which define a two-dimensional lattice for the diffraction orders. 
In particular, the target field information is encoded into the Fourier coefficient of the order $(m_1=-1, \; m_2=0)$ which is given by
\begin{align}
  T_{-1,0} = \frac{A}{\pi} e^{\im \phi},
\end{align}
where the linear relation between the height of the rectangles and $A$ can be appreciated. 
This linear dependence is achieved by creating a grating along the $\bs \nu_\perp$ direction which diverts energy from the filtered order into the transverse orders with $m_2\neq 0$.
This hologram offers the same diffraction efficiency as the parallel Lee hologram. 
Figure \ref{fig:ampcomp} also shows shaping of the same LG mode using this hologram. Here again, under appropriate conditions, the resulting shaped field is indistinguishable from the target, just like for the other holograms. 

It is worth noticing that for this hologram there are two carrier frequency vectors that need to be chosen. Nonetheless, similar rules apply to the choice of $\bs \nu_\perp$ as those that were mentioned earlier for $\bs \nu$. 
Here, $\norm{\bs \nu_\perp}$ controls the separation of the diffraction order along the $\bs \nu_\perp$ direction and so its norm should be larger that the bandwidth of the target field in order to sample it correctly and avoid overlap with the neighboring diffraction orders. 
Because of this, it is generally good practice to set  $\norm{\bs \nu_\perp} = \norm{\bs \nu}$, which is what will be assumed in the remainder of this work.
Nonetheless, for some particular fields for which the variations of amplitude and phase differ significantly, it is certainly possible to adapt these values.

\section{Discretization of the holograms}

\subsection{Lee holograms}

\begin{figure}
  \centering
  \includegraphics[scale=.75]{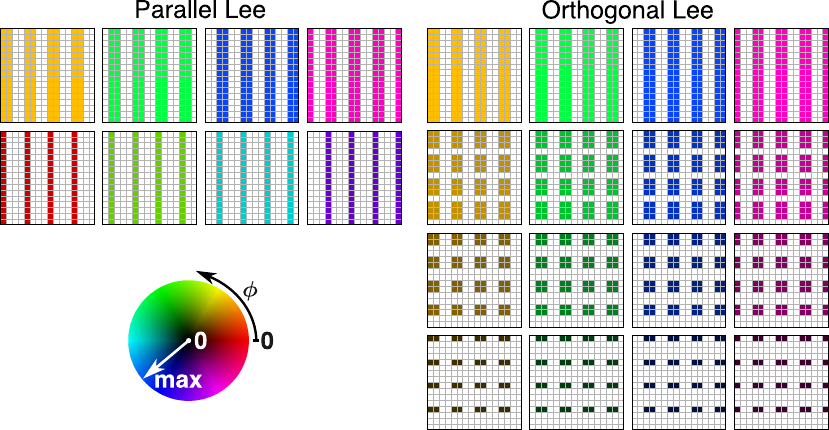}
  \caption{\label{fig:dis_cpx} 
  Discretization of the amplitude and phase values that can be encoded into the parallel and orthogonal Lee holograms when they are aligned with the Cartesian layout of the DMD's micro-mirrors when $\bs \nu = (1/4,0)$.
  The gratings are color coded according to the complex value they encode.}
\end{figure}

Up until now, we have been assuming that there are no restriction in the resolution a given hologram can have. 
But in reality, DMDs have a limited resolution dictated by the number of micro-mirrors which are assumed to be arranged on a Cartesian grid as shown in Fig.~\ref{fig:dis_cpx}. 
Given this Cartesian layout, one can argue that the most natural choice is to align the hologram with the pixel array with the carrier frequency pointing along the $x$ axis.
Taking into account the discretization induced by the micro-mirrors, in order to end up with a periodic grating, the carrier frequency values must be limited to the form $\bs \nu = (1/p, 0)$ where $p$ is an integer larger than one defining the periodicity of the grating in pixel units.
For simplicity, in everything that follows, all spatial units will be normalized by the pixel pitch defining the distance between neighboring pixels. 
In this way, all distances are in pixel units where one pixel is equal to one micro-mirror. 
A similar restriction holds for $\bs \nu_\perp$ in the orthogonal Lee hologram and so its value is set to $\bs \nu_\perp =  (0, 1/p)$.
It is also possible to write a single transmittance function encompassing both types of Lee holograms as 
\begin{align} \label{eq:leealrect}
  T(\bt r) =  &\sum_{n_1=-\infty}^{\infty}\sum_{n_2=-\infty}^{\infty} 
    \rect \left[ 
      \frac{ x-n_1 p -  s_x 
    }{w_x}
    \right] 
    \rect \left[ \frac{y - n_2 p 
    }{w_y}\right],
\end{align}
where $w_x$ and $w_y$ are the width and height of the rectangles, respectively, in pixel units, and $s_x$ controls the shift of the gratings. 
Its Fourier series is then given by 
\begin{align} \label{eq:leealrectfs}
  T(\bt r) = \sum_{m_1=-\infty}^{\infty}\sum_{m_2=-\infty}^{\infty} 
  \frac{\sin \left(\frac{\pi m_1 w_x}{p}\right)}{\pi m_1}
  \frac{\sin \left(\frac{\pi m_2 w_y}{p}\right)}{\pi m_2}
  e^{-2\pi \im m_1 \frac{s_x}{ p}}
    e^{\im  \frac{2\pi}{p} \left(m_1  x + m_2 y \right)},
\end{align}
from which the $(m_1=-1$, $m_2=0)$ diffraction order used to encode the target field can be seen to be given by
\begin{align} \label{eq:fcal}
  T_{-1,0} = \frac{w_y}{\pi p}\sin \left(\frac{\pi w_x}{p}\right)e^{2\pi \im \frac{s_x}{p}}.
\end{align}
Just as before, the spatial distribution of the target field is encoded into a combination of $w_x$, $w_y$, and $s_x$. 
In particular, since $w_x$ and $w_y$ can only take integer values between zero and $p$, from Eq.~(\ref{eq:fcal}) it can be seen that the discretization also affects the diffraction efficiency.
When $p$ is even the $\eta_d=1/\pi^2$ just like before, whereas when $p$ is odd $\eta_d = \sin^2[\pi(p-1)/2p]/\pi^2$ which is slightly lower but tends to $1/\pi^2$ as $p$ increases.

\begin{figure}
  \centering
  \includegraphics[scale=.75]{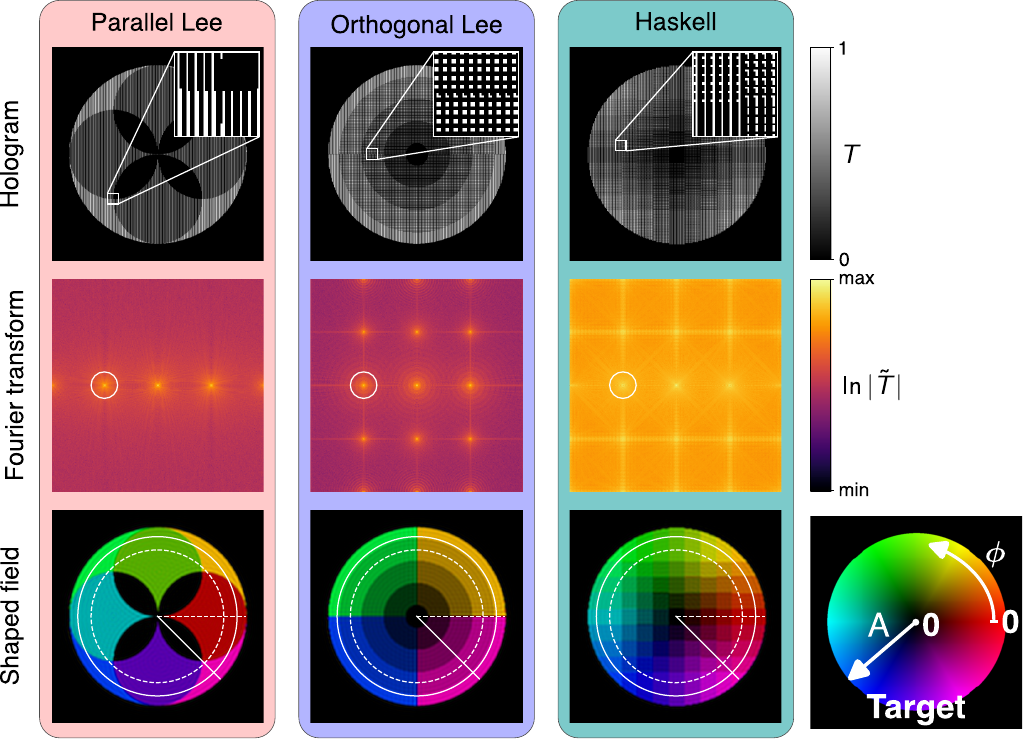}
  \caption{\label{fig:diskal} 
  Shaping of a target circular field with a vortex at the center and with a linearly increasing amplitude using the parallel and orthogonal Lee, and Haskell holograms aligned with the pixel array and with $\bs \nu = (1/4,0)$. 
  (first row) The binary holograms of size $800\times800$ aligned with the axis of the pixels, as shown in the inset. 
  (second row) Logarithm of the amplitude of the Fourier transform of the holograms where we can clearly see the various diffraction orders. The white circle denote the position and size of the aperture used for the filtering. 
  (third row) The shaped field where the variations along the overlaid white circles and segments are shown in Fig.~\ref{fig:amphial}.}
\end{figure}

Given the finite number of pixels within each period, $w_x$ and $w_y$ can take at most $p$ integer values which entails a quantization of the possible amplitude values that can be encoded in the hologram. 
However, when taking a closer look at the encoded amplitude dependence on $w_x$ and $w_y$ in Eq.~(\ref{eq:fcal}), it can be seen that there are only $\lfloor p/2 \rfloor$ values of $w_x$ leading to a different amplitude whereas there are $p$ values for $w_y$.
This is depicted in Fig.~\ref{fig:dis_cpx} for $p=4$, where the difference in the number of amplitude values that can be encoded between the parallel and orthogonal Lee holograms is evident. 
Likewise, for a given $w_x$ value $p$ distinct uniformly distributed phase values can be encoded. 
However, these values depend on the parity of the width $w_x$ since this it what dictates whether the center of the fringes coincide with the center of a pixel or with a boundary between two. 
Since in Eq.~(\ref{eq:leealrect}) it was assumed that the origin of the Cartesian axes lies in the middle of a pixel, when $w_x$ is odd then $s_x= 0,1,\ldots , p-1$, and when $w_x$ is even then $s_x=1/2, 3/2, \ldots, p-1/2$.
It must be noted that, in general, this supplementary phase values will be achieved for a different amplitude value thus creating a dependence between the amplitude and phase encodings for the parallel Lee hologram.
Therefore, for a given $w_x$ value $p$ distinct uniformly distributed phase values can be encoded. 
This quantization is again depicted in Fig.~\ref{fig:dis_cpx}.




\begin{figure}
  \centering
  \includegraphics[scale=.75]{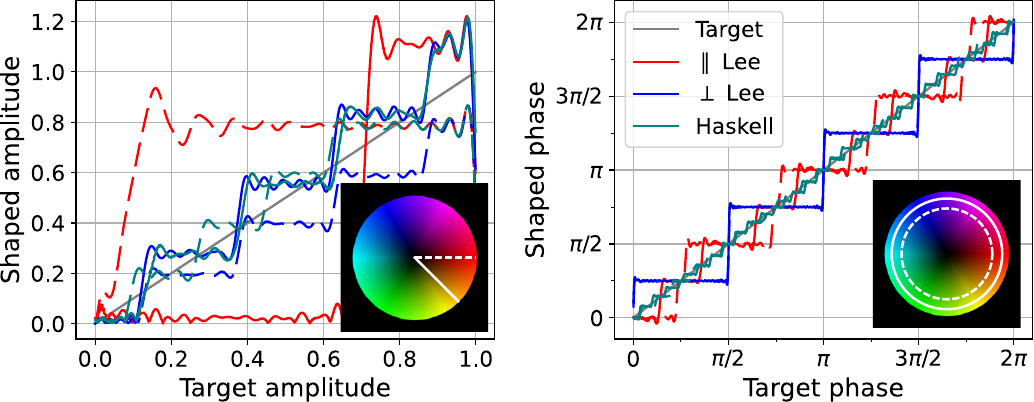}
  \caption{\label{fig:amphial} 
  (left) Amplitude and (right) phase variations obtained for the shaped field as a function of the target values along the white line segments and circles overlaid on the target field shown on the corresponding insets when using the Lee and Haskell holograms to produce the vortex disk field of Fig.~\ref{fig:diskal}.}
\end{figure}

To better understand the differences between the two types of Lee holograms, the shaping of a field with a phase vortex and with its amplitude increasing linearly from the center towards the edge is shown in Fig.~\ref{fig:diskal}. 
Here, regions where the phase and amplitude values are constant can clearly be appreciated.
This is a manifestation of the quantization of the attainable complex values due to the discretization of the micromirror layout. 
This quantization can be further appreciated in Fig.~\ref{fig:amphial} where the amplitude variation along different directions and the phase variations for varying radii exhibit a clear step-wise structure decorated by oscillations resulting from diffraction effects. 
Likewise, clear differences can be appreciated between the shaped fields resulting from the two types of holograms. 
The shaped field created by the parallel Lee holograms is not as uniform as the one created by the orthogonal one.
This is a consequence of encoding both the phase and amplitude values along the same direction which creates a dependence between the two and reduces the number of achievable amplitude values with respect to the orthogonal one. 
Likewise, the doubling of achievable phase values at different amplitude values for the parallel Lee can be appreciated.
Moreover, the nonlinear relationship relating the width to the amplitude for the parallel Lee concentrates most amplitude values around one.
In comparison, for the orthogonal Lee the achievable amplitude and phase modulations are independent. 
Only a slight difference is appreciable at the boundaries between phase regions which is due to the sudden change in the hologram pattern. 
All the corresponding gratings leading to these values are depicted in Fig.~\ref{fig:dis_cpx}.
It is worth emphasizing that, even though the expression in Eq.~(\ref{eq:leealrect}) has been used to gain a deeper understanding of the consequences of discretizing the holograms, in practice, the other expressions given in Eqs.~(\ref{eq:parlee}) and (\ref{eq:ortlee}) are used to generate the corresponding holograms since they are simpler to implement and computationally more efficient.



\begin{figure}
  \centering
  \includegraphics[scale=.75]{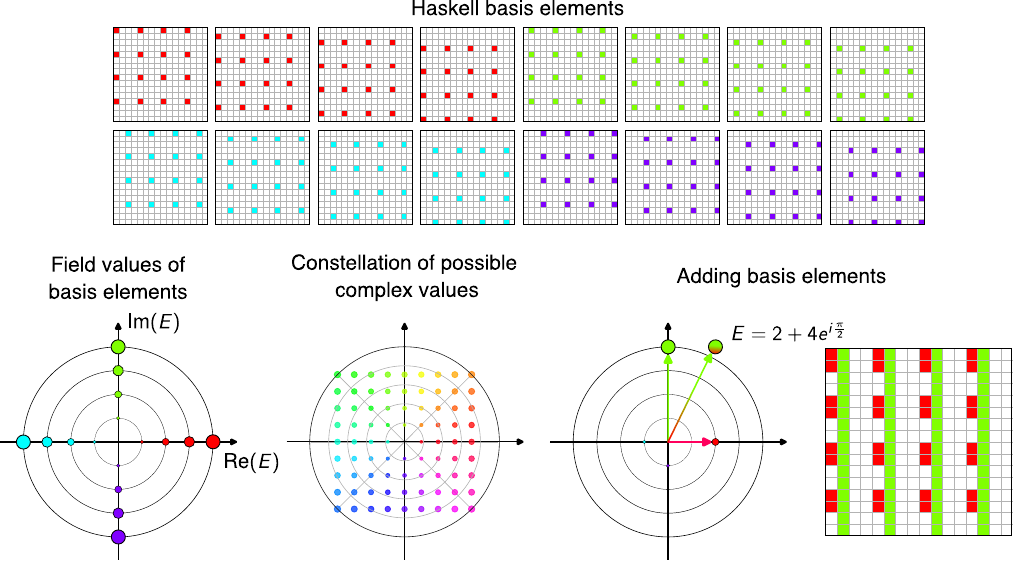}
  \caption{\label{fig:hask}
  Writing a binary Haskell hologram. The top figure shows all the possible grating basis elements when $p=4$ where the hue indicates the phase value that is encoded in them. The several basis elements which encode the same phase value can be combined to encode higher amplitude values that make up the various field values of basis elements.
  These field values can then be combined in order to obtain the constellation of complex values shown in the bottom row. 
  To encode any such value a look-up table is used to determine the appropriate combination of basis elements and grating structure.}
\end{figure}

Despite the differences between both types of Lee holograms, neither of them produces a satisfactory shaped field in Fig.~\ref{fig:diskal}. 
In both cases, large areas with constant amplitude and phase values are clearly visible thus causing significant deviations from the target field.
This problem arises from the limited number of complex values that can be encoded into the hologram as shown in Fig.~\ref{fig:dis_cpx}. 
A possible solution would be to increase $p$ since the number of both amplitude and phase values increases with $p$.
The problem with this approach is that now more pixels of the DMD need to be used to encode a specific complex value. 
Assuming the target field varies smoothly, at least $p\times p $ sized super pixels need to be used to appropriately encode a given complex value. 
The number of super pixels in essence represents the number of available degrees of freedom on the DMD that can be used to shape light and thus limits the bandwidth of the fields that can be generated. 
Therefore, the ideal solution is to keep the same $p$ while finding alternative ways to increase the number of achievable phase and amplitude values. 

\subsection{Haskell hologram}

One possible solution to increase the number of complex values that can be encoded while keeping the same value for $p$ is to merge both types of Lee holograms.
In this case, the hologram would be composed of rectangles whose position, width and height would be adjusted to encode a target complex value. 
However, since there is no univocal relation between a complex value and the parameters of the rectangles, i.e. there can be several rectangles that encode the same complex value, a look-up table mapping a complex value to a specific rectangle would have to be constructed. 
However, if a look-up table is going to be used, there is a better alternative. 
As proposed by Haskell more than half a century ago \cite{haskell1973computer,fienup1974new}, one can instead consider combinations made out of the smallest units used to encode a hologram with a given period. 
These unit elements can then be used as basis elements to create more general grating structures that need not be composed of rectangles but can have more general shapes which can lead to supplementary phase and amplitude values. 
Figure~\ref{fig:hask} shows the basic principle behind this method.  

A Haskell hologram is a two-dimensional array with primitive vectors $\bt b_1 = (p,0)$ and $\bt b_2 = (0,p)$ of superpixels formed by grouping $p \times p$ pixels.
As shown in Fig.~\ref{fig:hask}, the Haskell basis elements are those in which a single pixel within the superpixel is turned on and thus form  
rectangular gratings with $w_x=w_y=1$, and varying shifts along $x$ and $y$.
A given target complex value is encoded by finding the superposition of these basis elements that encode the complex value that is closest to the target. 
To determine which complex values can be encoded into these holograms, it is necessary to derive the Fourier coefficient of the $(m_1=-1, m_2=0)$ diffraction order of the transmittance function obtained from such a superposition. 
From Eqs.~(\ref{eq:leealrect}) and (\ref{eq:leealrectfs}) it can be deduced to be given by,
\begin{align} 
  T_{-1,0} = \sum_{s_x=1}^{p}\sum_{s_y=1}^{p} 
    \frac{c_{s_xs_y}}{\pi p}\sin \left(\frac{\pi}{p}\right)e^{2\pi \im \frac{s_x-1 }{p}},
\end{align}
where the coefficients $c_{s_xs_y}$ are either one or zero. 
From this expression, it can be seen that the basis elements that differ in a shift $s_x$ along $x$ encode different phase values. 
This leads to the encoding of $p$ distinct phase values that are uniformly distributed between $0$ and $2\pi$ and with the same amplitude encoding. 
From the previous equation it can be seen that there is a clear degeneracy along the $y$ direction since the $p$ basis elements corresponding to different shifts $s_y$ lead to the same complex value encoding as depicted in Fig.~\ref{fig:hask}.
Additionally, it is generally possible to have different combination of basis elements with different phase values that encode the same complex value.
Therefore, a look-up table mapping a complex value to a specific combination of the basis elements needs to be created since there is no univocal relationship between the two. 
In practice, when constructing a Haskell hologram, the first step is to take the image of the target complex field having the same resolution as that of the DMD and down sampling it by a factor of $p$ along each direction.
In this way each point of the down-sampled array identifies the complex value that needs to be encoded into the corresponding superpixel.
The value of each pixel within the superpixel is determined through the look-up table.  

To get a better sense of which complex values can be encoded into a Haskell hologram  it is possible to remove the shift $s_y$ from the Fourier coefficient by writing it as
\begin{align} \label{eq:alignfc}
  T_{-1,0} = \sum_{s_x=1}^{p}
    \frac{C_{s_x}}{\pi p}\sin \left(\frac{\pi}{p}\right)e^{2\pi \im \frac{s_x-1 }{p}},
\end{align}
where $C_{s_x}$ can take integer values from zero to $p$, and simply represents the number of pixels along $y$ that are set to one, but does not specify which ones. 
In this way, the different field values that can be encoded with each basis elements correspond to those with the same phase value as before but with $p$ amplitudes values uniformly distributed between $0$ and $p$.
The constellation of possible complex values that can be encoded is then determined by considering all possible combinations of the basis elements as shown in Fig.~\ref{fig:hask}.

Figures \ref{fig:diskal} and \ref{fig:amphial} show the results of using the Haskell hologram to shape the same field that was used to study the discretization effects on the Lee holograms. 
The improvements are evident, while there are still regions with constant phase and amplitude values, they are much smaller and thus produce a much smoother shaped field that clearly resembles the target field. 
This smoothness is particularly visible in Fig.~\ref{fig:amphial} for the phase modulations which deviate only slightly from the target values. The amplitude values, however, could still be improved since the achievable values remain noticeably depend on the particular direction. 
Now, to readers that might be familiar with the different holograms that have been proposed to shape light with DMDs, this method might seem equivalent to the superpixel method proposed more recently in Ref.~\cite{goorden2014superpixel}. 
However, they differ in a small detail that has a big impact on the number of complex values that can be encoded. Nonetheless, Haskell's holograms can be seen as the direct ancestor to the superpixel hologram even though this connection appears to have been lost over time.

\begin{figure}
  \centering
  \includegraphics[scale=0.75]{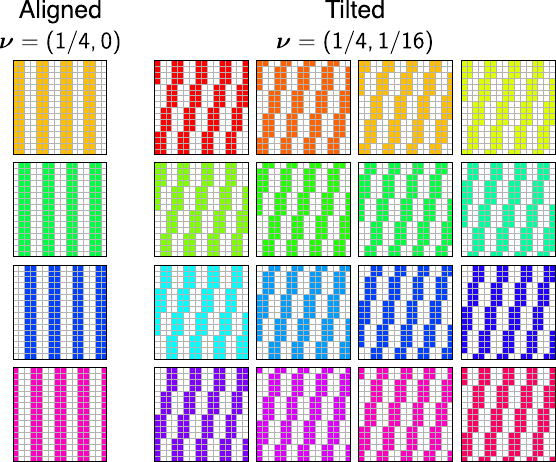}
  \caption{\label{fig:phases} 
  Phase values that can be encoded into a binary hologram when it is aligned with $\bs \nu= (1/4,0)$ and when it is tilted with $\bs \nu = (1/4,-1/16)$. 
  The drastic increase of phase values enabled by tilting the hologram is clearly appreciable.}
\end{figure}

\section{Tilting the hologram}

\subsection{Lee holograms}

While aligning the hologram with the pixel array seems natural, it might not be the best choice.
Let us then consider what happens when the hologram is tilted with respect to the axes of the pixel array. 
For the sake of simplicity, only the particular case using of the holograms for phase modulation will be treated in full. 
Given the discretization of the pixels, in order to get a periodic structure in the hologram, the components of the carrier frequency vector need to be chosen as $\bs \nu  = (1/p_x,-1/p_y)$ where $p_y/p_x = k$ with $k$ being an integer. 
Here it was assumed that $p_y\geq p_x$ without loss of generality. 
Given these assumptions, the resulting hologram is formed by a two-dimensional lattice of rectangles of width $w_x = \lfloor p_x /2\rfloor$ and height $w_y=k$ and whose primitive vectors are $\bt a_1= (p_x,0)$ and $\bt a_2 = (1,k)$. 
Examples of this lattice are shown in Fig.~\ref{fig:phases} for $p_x =1/4$ and $p_y=1/16$.

The transmittance function for this hologram can be written as
\begin{align} \label{eq:tiltrect}
  T(\bt r) =  &\sum_{m_1=-\infty}^{\infty}\sum_{m_2=-\infty}^{\infty} 
    \rect \left[ \frac{ x-m_2 p_x - m_1 - s_x}{w_x}\right] 
    \rect \left[ \frac{y -k m_1 - s_y}{w_y}\right].
\end{align}
As before, in order to understand which phase values can be encoded into this hologram, the transmittance function is expanded into a Fourier series giving
\begin{multline} \label{eq:tiltfs}
  T(\bt r) = 
  \sum_{m_1=-\infty}^{\infty}
  \sum_{m_2=-\infty}^{\infty} \frac{\sin \left( \pi m_1\frac{w_x}{p_x} \right)}{\pi m_1}\frac{\sin \left[\pi \left(m_2-\frac{m_1}{p_x}\right) \frac{w_y}{k}\right]}{\pi \left(m_2-\frac{m_1}{p_x}\right)}
  e^{-\im 2 \pi m_1 (\frac{s_x}{p_x}-\frac{s_y}{p_y})}
  e^{-\im 2 \pi m_2 \frac{s_y}{k}}
  \\
  \times
  e^{\im 2 \pi m_1 (\frac{x}{p_x}-\frac{y}{p_y})}
  e^{\im 2 \pi m_2 \frac{y}{k}}.
\end{multline}
Here, it can be seen that the diffraction pattern creates a reciprocal two-dimensional lattice with $\bt b_1=\bs \nu =(1/p_x,-1/p_y)$ and $\bt b_2 = (0,1/k)$ which satisfy $\bt a_i \cdot \bt b_j = \delta_{ij}$.
By filtering the $(m_1=-1, m_2=0)$ order, the shaped field will be proportional to 
\begin{align} \label{eq:tiltfc}
  T_{-1,0} = 
  \frac{p_x}{\pi^2}
  \sin \left( \pi \frac{w_x}{p_x} \right)\sin \left(\pi \frac{ w_y}{p_x k}\right)
  e^{\im 2 \pi  (\frac{s_x}{p_x}-\frac{s_y}{p_y})}.
\end{align}
The first thing to notice from this expression is the total number of possible phase values that can be encoded. 
The smallest phase shift is now $2\pi/p_y$ which is produced when shifting the hologram by one pixel along the $y$ direction. 
Therefore, the total number of phase values that can be encoded for a given value of $w_x$ and $w_y$ is increased by a factor of $k$. 
This drastic increase in phase values is exemplified in Fig.~\ref{fig:phases}. 

\begin{figure}
  \centering
  \includegraphics[scale=0.75]{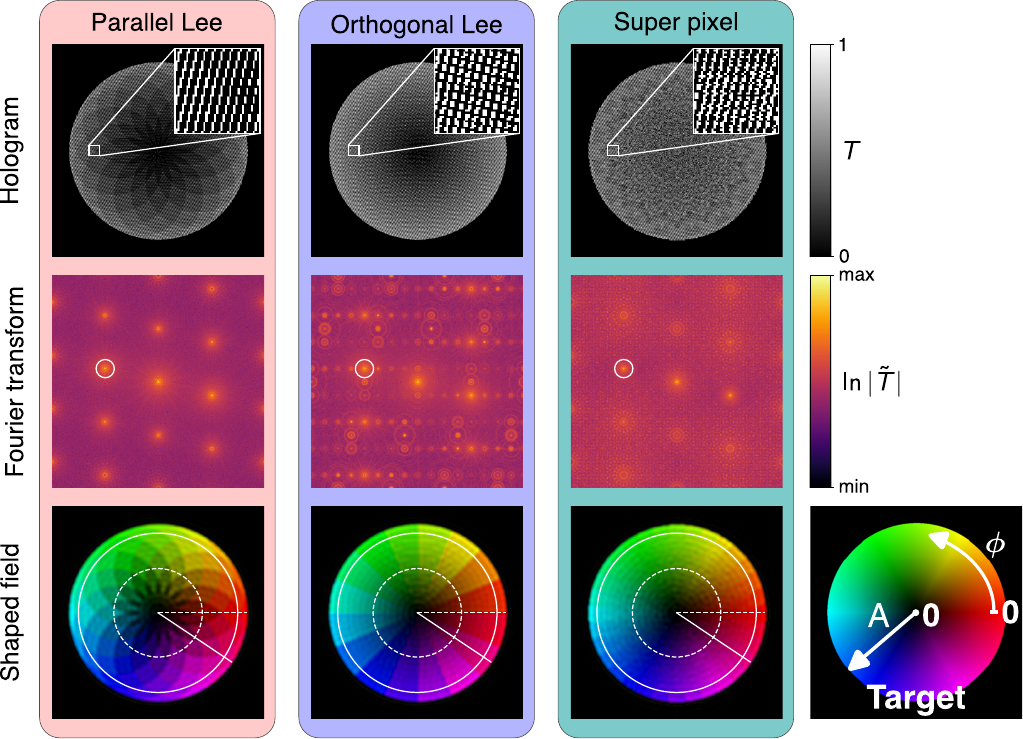}
  \caption{\label{fig:diskang} 
  Shaping of a target circular field with a vortex at the center and with a linearly increasing amplitude using the parallel and orthogonal Lee, and superpixel holograms tilted with respect to the pixel array with $\bs \nu = (1/4,-1/16)$. 
  (first row) The binary holograms of size $800\times800$ aligned with the axis of the pixels, as shown in the inset. 
  (second row) Logarithm of the amplitude of the Fourier transform of the holograms where we can clearly see the various diffraction orders. The white circle denote the position and size of the aperture used for the filtering. 
  (third row) The shaped field where the variations along the overlaid white circles and segments are shown in Fig.~\ref{fig:amphial}.}
\end{figure}

A crucial point is that this increase in the number of phase values comes at no cost in the number of degrees of freedom of the DMD, and thus of the resolution of the shaped field, with respect to the hologram obtained when $\bs \nu= (1/p_x,0)$ as long as $ k\leq p_x$.
The only significant cost comes from the smaller prefactors in Eq.~(\ref{eq:tiltfc}) when compared to those in Eq.~(\ref{eq:alignfc}) which diminish the diffraction  efficiency of the grating since some light is now being diffracted along the $y$ direction. 
This might be an important factor to consider if minimizing losses is more important than increasing the shaping accuracy. 
It is also important to determine the optimal tilt angle of the hologram which is determined by the choice of $p_y$ for a given $p_x$.
From Eq.~(\ref{eq:tiltfc}) it can be seen that if the hologram is tilted at 45$^{\circ}$, which is one of the most common choices, then $p_y=p_x$ and $k=1$ so that there is no increase in the number of phases values that can be encoded. 
In contrast, the smaller the tilt the larger $p_y$ and $k$ are, and so is the number of phase values. 
However, if $p_y>p_x^2$ then the unit cell of the lattice starts becoming larger which lowers the resolution with respect to the aligned hologram. 
A good compromise is to take $p_y=p_x^2$. 

\begin{figure}
  \centering
  \includegraphics[scale=0.75]{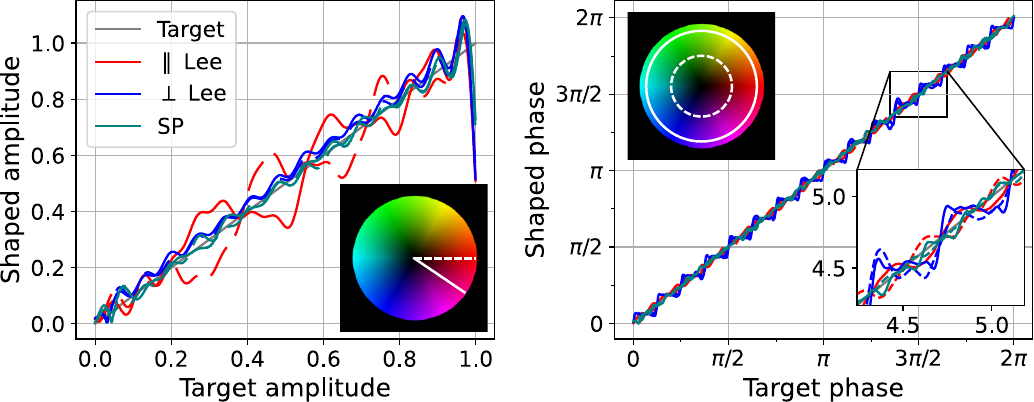}
  \caption{\label{fig:amphiang} 
  (left) Amplitude and (right) phase variations obtained for the shaped field as a function of the target values along the white line segments and circles overlaid on the target field shown on the corresponding insets when using the Lee and super pixel holograms to produce the vortex disk field of Fig.~\ref{fig:diskang}.}
\end{figure}

Figures \ref{fig:diskang} and \ref{fig:amphiang} show the drastic improvement provided by tilting the hologram on the shaping of the same target field as the one used in Figs.~\ref{fig:diskal} and \ref{fig:amphial}. 
For both Lee holograms it can be seen that the phase variations are much smoother, with the step-like dependence being barely noticeable in Fig.~\ref{fig:amphiang}. 
Nonetheless, there is a clear difference for the amplitude variation between the two holograms.
For the orthogonal Lee hologram the amplitude variations are fairly smooth whereas for the parallel Lee hologram large deviations from the target values can still be observed which depend on the phase value being encoded. 
Therefore, as was mentioned before, decoupling the phase and amplitude encodings allows for the independent encoding of the amplitude and phase information while also creating more uniform and smoother fields.

\subsection{Superpixel hologram}

\begin{figure*}
  \centering
  \includegraphics[width=.95\linewidth]{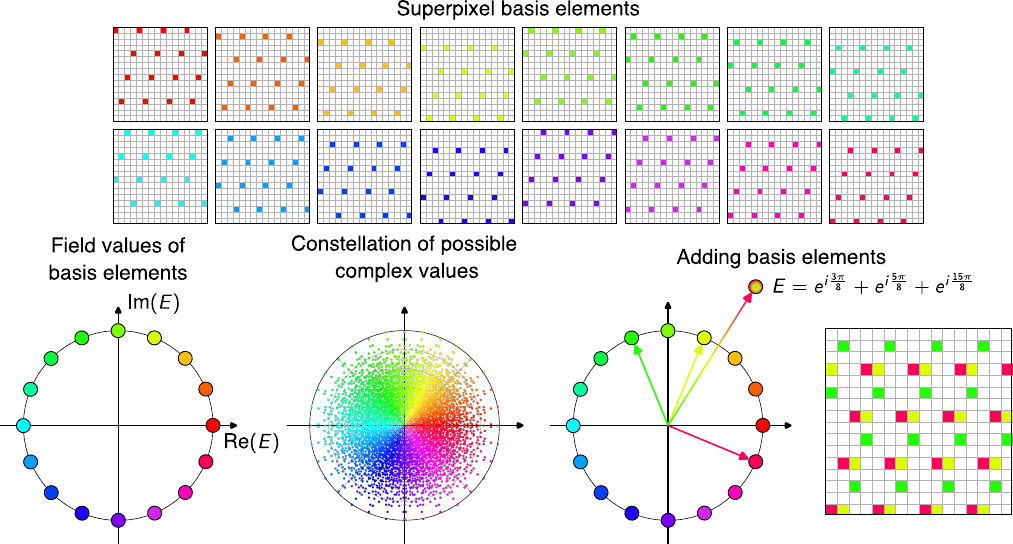}
  \caption{\label{fig:sp} 
  Writing a binary Haskell hologram. The top figure shows all the possible grating basis elements when $p=4$ where the hue indicates the phase value that is encoded in them. The several basis elements which encode the same phase value can be combined to encode higher amplitude values that make up the various field values of basis elements.
  These field values can then be combined in order to obtain the constellation of complex values shown in the bottom row. 
  To encode any such value a look-up table is used to determine the appropriate combination of basis elements and grating structure.}
\end{figure*}

Even though tilting the Lee holograms allows for a much smoother an accurate shaping of light, some applications might require a better field accuracy.
As for the case in which hologram was aligned with the micro-mirror array, the ideal scenario would be to achieve a higher number of complex values without altering the resolution, that is keeping $\bs \nu$ constant. 
Taking a similar approach as for the Haskell hologram, it is possible to construct a more general type of hologram by superimposing the smallest grating units that satisfy the same lattice structure, and thus the same primitive vectors. 
For given $p_x$ and $p_y$, there are a total of $k\times p_x = p_y$ distinct basis elements with their transmittance function being equal to that in Eqs.~(\ref{eq:tiltrect}) and (\ref{eq:tiltfs}) with $w_x=w_y=1$. 
Therefore, the Fourier coefficient for the $(m_1=-1,m_2=0)$ diffraction order resulting from a superposition of the various basis elements can be directly written as
\begin{align}
  T_{-1,0} = \frac{p_x}{\pi^2}
  \sin \left(  \frac{\pi}{p_x} \right)\sin \left(\frac{\pi}{p_y}\right)
  \sum_{s_x=1}^{p_x}\sum_{s_y=1}^k  c_{s_xs_y}  
  e^{\im 2 \pi  \left(\frac{s_x}{p_x}-\frac{s_y}{p_y}\right)},
\end{align}
where the coefficients $c_{s_xs_y}$ are either one or zero. 
Therefore, this superpixel hologram can encode any complex value that results from the superposition of $p_x \times k$ complex numbers uniformly distributed along a circle in the complex plane~\cite{putten2008spatial,goorden2014superpixel}. 
Figure \ref{fig:sp} shows all the basis elements for the particular case $p_x=4$ and $p_y=16$, along with the resulting complex values and an example the resulting diffraction grating for encoding a particular complex value.
This type of hologram goes by the name of superpixel, since a given target complex value is encoded in a superpixel made of $p_x \times k$ pixels whose values are chosen with a look-up table to determine the appropriate combination that provides the closest value to the target. 
The enhanced shaping capabilities provided by the superpixel method can be fully appreciated in Figs.~\ref{fig:diskang} and \ref{fig:amphiang}. 
It provides both the smoothest phase and amplitude variations. 
It must be noted that most amplitude oscillations come from diffraction effects stemming from the aperture used for the filtering and not the discretization of amplitude values in the hologram. 

Having introduced Haskell's method, it is clear that the only difference, albeit a crucial one, is the tilt that is introduced for the superpixel method which allows lifting the degeneracy for the phase values. 
Note that for the original superpixel method $p_y=p_x^2$ which is the smallest value of $p_y$ which fully lifts the phase degeneracy and creates square superpixels. 
While not stressed in the original work \cite{goorden2014superpixel}, this tilt is a crucial component for the superpixel method. 
One cannot simply treat each superpixel independently since neighboring superpixels along $y$ need to be shifted by one pixel along $x$ in order to create the appropriate hologram. 
This shift can be seen in Fig.~\ref{fig:sp}.

\section{Closing remarks}

\begin{figure*}
  \centering
  \includegraphics[width=.95\linewidth]{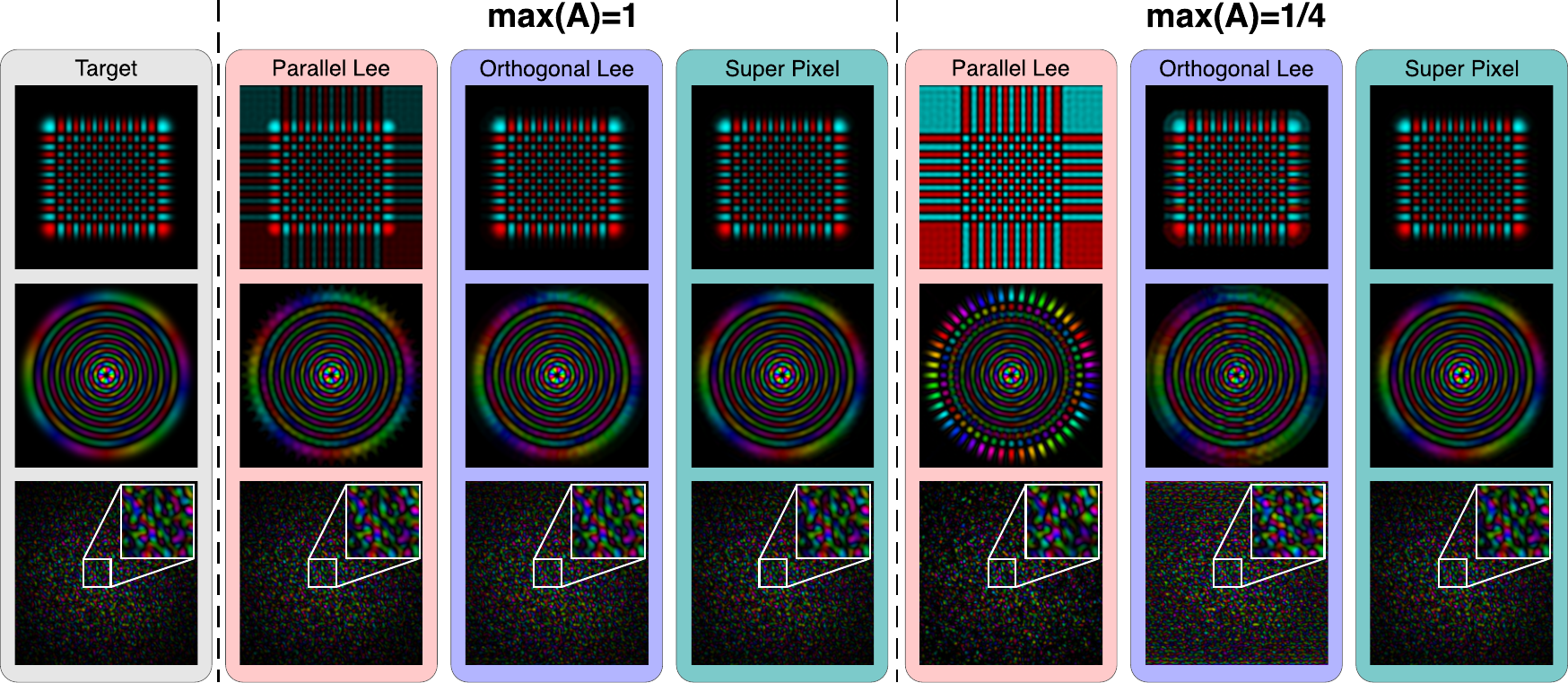}
  \caption{\label{fig:cpx} 
  Shaping of complex optical fields using tilted Lee and superpixel holograms with $\bs \nu = (1/4,-1/16)$ for two maximum values for the amplitude modulation. 
  The complex fields being considered are high-order (first row) Hermite-Gauss (HG) and (second row) LG beams and (third row) a speckle field obtained by from the random superposition of 100 plane waves with a Gaussian envelope.
  The holograms used to generate the shaped fields, as well as all the fields that are shown have a size of $780\times780$ pixels. }
\end{figure*}

Despite DMDs only providing a binary amplitude modulation with a finite number of pixels, it is possible to shape complicated light fields through specially designed holograms, which include the parallel and orthogonal Lee holograms, and the superpixel method. 
This is exemplified in Fig.~\ref{fig:cpx}, where these three types of holograms have been used to successfully shape higher-order Hermite-Gauss (HG) and LG beams, as well as a speckle field.
This, however, is only achieved by appropriately tilting the hologram with respect to the axes of the micro-mirror array so that the carrier frequency of the hologram is given by $\bs \nu =(1/p_x, 1/p_x^2)$, with $p_x$ being the period along the $x$ direction of the array in pixel units. 
This choice allows for keeping the same number of degrees of freedom as those of the holograms that are aligned with $\bs \nu =(1/p_x, 0)$ or at 45${}^\circ$ with $\bs \nu =(1/p_x, 1/p_x)$, but drastically increases the number of complex values that can be encoded. 
The improvement provided by this optimal tilt is evident from the field and intensity correlation values shown in Table~\ref{tab} for various values of the period $p_x$, particularly for the parallel Lee hologram. 
For each field, the size of the aperture was chosen to be slightly larger than their bandwidth. 
It should be mentioned that the correlations resulting form the hologram at 45${}^\circ$ are almost identical to those obtained with the aligned holograms. 
Because of this they have been omitted from Table~\ref{tab}.

Taking a closer look at the correlation values in Table~\ref{tab}, it is possible to draw some general conclusion. 
Firstly, as $p_x$ increases, the difference between the tilted and aligned holograms goes away for all except the parallel Lee hologram, which seems to always benefit from the tilt due to its more restricted phase modulation. 
However, as $p_x$ increases it reaches a point in which the correlation starts diminishing since the lower number of degrees of freedom available are not sufficient to properly resolve fields with larger bandwidths. 
This effect can be appreciated for the speckle field, which is the one with the largest bandwidth. 
It is also noteworthy that the superpixels methods do not always produce a more accurately shaped field than the corresponding tilted Lee holograms.
Alongside the correlation values, the diffraction efficiencies for each type of hologram are also shown in Table~\ref{tab}. 
These values show the drop that is caused by tilting the hologram, and that holograms with even $p_x$ have better efficiencies than the neighboring odd values.

Large differences between the shaped fields generated by the different types of tilted holograms start to appear when the amplitude modulation is limited, that is, when the maximum amplitude of a given field is set to a value lower than one. 
This limited modulation can appear when one uses the DMD to regulate the total input power, or when fixing the total amount of energy on the input fields so that the field with the highest amplitude enjoys the full amplitude modulation, but the amplitude modulation for all others is limited. 
The effect of limiting the amplitude modulation is shown in Fig.~\ref{fig:cpx} for the same fields as before. 
There, it can be seen that the parallel Lee hologram presents the largest deviations from the target field, which was to be expected. 
As mentioned in previous sections, coupling the phase and amplitude encoding into the hologram along the same direction limits the number of achievable amplitude values.
Moreover, it creates a dependency between the amplitude and phase encodings, which leads to the angular lobes in the shaping of the LG beam. 
For the orthogonal Lee holograms the resulting fields are noticeably closer to the targets, but some differences are clearly visible. 
In contrast, for the superpixel method there is no appreciable difference in the shaped field. 
This is due to the large number of complex values that can be encoded through this method, with many of them having an amplitude that is lower than the maximal one, as shown in Fig.~\ref{fig:sp}.

\newcolumntype{N}{@{}m{0pt}@{}}
\newcolumntype{M}[1]{>{\centering\arraybackslash}p{#1}}
\newcolumntype{F}{>{\centering\arraybackslash}m{.8cm}}
\begin{table}
  \caption{\label{tab}
    Field ($C_E$) and intensity ($C_I$) correlations between the shaped and target fields. 
  The target fields are the same as those shown in Fig.~\ref{fig:cpx} and the shaped fields were obtained using aligned and tilted versions of the parallel ($\parallel$) and orthogonal ($\perp$) Lee holograms for four different values of the period. 
  For $p_x=3$ and $4$ both the results of the Haskell and super pixel holograms were also included.}
\begin{center}
  \begin{tabular}{|c| c||M{1.6cm}|
    F|F|F|F|F|F|
    N|}
    \hline
    \rule{0pt}{3ex} 
    \multirow{2}{1.1cm}{\centering Period ($p_x$)} & 
    \multirow{2}{*}{Hologram} & 
    \multirow{2}{1.6cm}{\centering 
    $\eta_d$ (\%)}&
    \multicolumn{2}{c|}{HG beam} & 
    \multicolumn{2}{c|}{LG beam} & 
    \multicolumn{2}{c}{Speckle}&\\[2.5pt]
    \cline{4-6}\cline{7-9}
    \rule{0pt}{3ex} 
    & & & 
    $C_E$ & $C_I$ & $C_E$ & $C_I$ & $C_E$ & $C_I$ & \\[2.5pt]
    \hline \hline
    \multirow{6}{*}{3} 
    & Aligned $\parallel$ Lee &7.6&43.6 & 13.8 & 60.2 & 36.7 & 73.4 & 55.9 & \\
& Aligned $\perp$ Lee &7.6&99.6 & 99.6 & 96.4 & 96.9 & 93.6 & 95.4 & \\
& Tilted $\parallel$ Lee ($k=p_x$) &5.2&77.0 & 85.1 & 94.8 & 91.7 & 95.7 & 94.9 & \\
& Tilted $\perp$ Lee ($k=p_x$) &5.2&99.7 & 99.8 & 99.6 & 99.7 & 96.2 & 96.6 & \\
& Haskell &7.6&99.3 & 99.3 & 96.5 & 97.2 & 92.1 & 94.5 & \\
& Super pixel &5.2&99.8 & 99.9 & 99.7 & 99.7 & 96.6 & 96.6 & \\
    \hline    
    \multirow{6}{*}{4} 
    & Aligned $\parallel$ Lee &10.1&64.3 & 39.7 & 68.9 & 47.6 & 82.1 & 68.4 & \\
& Aligned $\perp$ Lee &10.1&99.7 & 99.7 & 92.3 & 95.5 & 93.8 & 94.4 & \\
& Tilted $\parallel$ Lee ($k=p_x$) &8.2&87.1 & 97.7 & 99.4 & 99.3 & 96.5 & 96.1 & \\
& Tilted $\perp$ Lee ($k=p_x$) &8.2&99.6 & 99.7 & 99.4 & 99.3 & 94.9 & 95.4 & \\
& Haskell &10.1&99.5 & 98.3 & 99.1 & 98.0 & 95.8 & 95.4 & \\
& Super pixel &8.2&99.7 & 99.9 & 99.8 & 99.8 & 94.8 & 95.2 & \\
    \hline    
    \multirow{4}{*}{5} 
    & Aligned $\parallel$ Lee &9.2&52.6 & 36.1 & 76.2 & 59.1 & 86.9 & 78.1 & \\
& Aligned $\perp$ Lee &9.2&99.7 & 99.8 & 99.0 & 99.4 & 96.6 & 96.7 & \\
& Tilted $\parallel$ Lee ($k=p_x$) &8.0&96.4 & 99.5 & 99.9 & 99.8 & 96.4 & 95.5 & \\
& Tilted $\perp$ Lee ($k=p_x$) &8.0&99.7 & 99.9 & 99.6 & 99.6 & 91.9 & 93.2 & \\
    \hline    
    \multirow{4}{*}{6} 
    & Aligned $\parallel$ Lee &10.1&65.2 & 46.9 & 83.0 & 69.2 & 89.1 & 83.3 & \\
& Aligned $\perp$ Lee &10.1&99.8 & 99.9 & 97.1 & 98.6 & 96.2 & 96.5 & \\
& Tilted $\parallel$ Lee ($k=p_x$) &9.2&97.1 & 99.8 & 99.9 & 99.8 & 95.7 & 94.2 & \\
& Tilted $\perp$ Lee ($k=p_x$) &9.2&98.1 & 99.0 & 97.7 & 97.6 & 93.8 & 94.2 & \\
    \hline                    
  \end{tabular}
\end{center}
\end{table}

Taking all these considerations into account, choosing $\bs \nu =(1/p_x,-1/p_x^2)$ with $p_x=4$ provides a good compromise between the number of complex values that can be encoded and the number of degrees of freedom.
This choice allows for the accurate shaping of complex field with large bandwidth, i.e. fast spatial variation.
While for most cases the tilted Lee holograms should be sufficient, this choice of $\bs nu$ also allows using the super pixel method if accurate modulation is required for different amounts of energy. 
Moreover, all the codes needed to implement these holograms can be found in the dedicated GitHub repository along with examples of use~\cite{gutierrezcuevaspyDMDholo}. 



As a last note it must be mentioned that there are other methods that have been used to shape light using a DMDs. 
For example, it is possible to use their rapid switching to modulate the amplitude detected by a camera with a large integration time. 
However, this type of amplitude modulation produces incoherent light which might not be suitable for many applications.
There have been, nonetheless, other proposals for the shaping of coherent lights. 
In Refs.~\cite{ulusoy2011full,zlokazov2020methods}, it was suggested to add additional masks at the Fourier plane in order to tailor the number and distribution of complex numbers that can be encoded using a DMD.
While this approach also opens the door to using DMDs as information processing systems, its implementation is experimentally more complicated and unnecessary for most cases. 
There are only two other options for writing binary holograms to shape coherent light and that use the optical system shown in Fig.~\ref{fig:setup}.
The first is based on using an iterative method to design a hologram that produces a target intensity or field distribution \cite{lenton2020otslm}. 
However, this method is more computationally expensive than the holograms presented here and with no guarantee that it will yield a better shaped field. 
Therefore, if the desired target field distribution is known it is best to use one of the holograms presented here. 
The second technique is based on using pixel dithering to convert the grayscale hologram of Eq.~(\ref{eq:oaholo}) into a binary one \cite{lerner2012shaping,cheremkhin2019comparative}. 
In this case the phase is encoded in the position of the fringes of the grating and the amplitude is modulated by setting specific pixels to zero or one in order to mimic the grayscale variation of the original hologram. 
The resulting hologram is then quite similar to that obtained with the orthogonal Lee hologram but is computationally more expensive and harder to implement. 
In which case, one would be better off using the superpixel method.

\section*{Acknowledgements}

R.G.C. also acknowledges funding from the Labex WIFI 
(ANR-10-LABX-24, ANR-10-IDEX-0001-02 PSL*).

\section*{Disclosures}

The authors declare no conflicts of interest.

\section*{Data Availability Statement}

The code used to generate all the holograms presented in this work and to simulate the shaping of complex fields is available in Ref.~\cite{gutierrezcuevaspyDMDholo}.

\bibliography{DMDtuto}

\end{document}